\documentclass{osa-article}

\journal{osajournal}

\articletype{Research Article}

\begin{document}

\title{Entropic uncertainty relations and the measurement range problem, with consequences for high-dimensional quantum key distribution}

\author{J. Eli Bourassa\authormark{1,*}, and Hoi-Kwong Lo\authormark{1,2,3}}

\address{\authormark{1}Department of Physics, University of Toronto, 60 St. George St., Toronto, Ontario, M5S 1A7, Canada

\authormark{2}Center for Quantum Information and Quantum Control, University of Toronto, 80 St. George St., Toronto, Ontario, M5S3H6, Canada

\authormark{3}Department of Electrical and Computer Engineering, University of Toronto, 10 King's College Road, Toronto, Ontario, M5S 3G4, Canada\\}

\email{\authormark{*}bourassa@physics.utoronto.ca} 

\begin{abstract}
The measurement range problem, where one cannot determine the data outside the range of the detector, limits the characterization of entanglement in high-dimensional quantum systems when employing, among other tools from information theory, the entropic uncertainty relations. Practically, the measurement range problem weakens the security of entanglement-based large-alphabet quantum key distribution (QKD) employing degrees of freedom including time-frequency or electric field quadrature. We present a modified entropic uncertainty relation that circumvents the measurement range problem under certain conditions, and apply it to well-known QKD protocols. For time-frequency QKD, although our bound is an improvement, we find that high channel loss poses a problem for its feasibility. In homodyne-based continuous variable QKD, we find our bound provides a quantitative way to monitor for saturation attacks.
\end{abstract}

\section{Introduction}
Entropic uncertainty relations have proven to be powerful tools for quantum information, as they provide a method for quantitatively relating the statistics of measurement outcomes, the operators characterizing those measurements, and the amount of information different parties can have about a system \cite{Maassen1988,Berta2010,Konig2009,Tomamichel2011,Tomamichel2012a,Coles2012,Coles2014a,Coles2017}. This proves particularly useful for determining the security offered by quantum key distribution (QKD) protocols, in which trusted parties, Alice and Bob, must bound the information held by an eavesdropper, Eve \cite{Renner2005,Leverrier2010,Furrer2012,Tomamichel2012,Nunn2013,Niu2016}. In particular, in an entanglement-based protocol, the optical source and channels may not be trusted, so Alice and Bob must rely on the characterization of their measurements and the statistics of their outcomes to determine security \cite{Bennett1992}. 

Fundamental bounds on the secure key rate exist for a point-to-point quantum channel \cite{Pirandola2017,Takeoka2014}. Nonetheless, there has been recent interest in creating more efficient QKD strategies that employ high-dimensional photonic degrees of freedom to maximize secret bits per optical signal. In particular, time-energy \cite{Qi2006,Qi2011,Nunn2013,Lee2014,Zhong2015,Islam2017a,Islam2017,Brougham2013,Bunandar2015,Niu2016,Mower2013,Zhang2014}, orbital angular momentum \cite{Sit2016}, and electric field quadrature \cite{Grosshans2003,Furrer2012,Laudenbach2018,Gehring2015} are all candidates for high-dimensional degrees of freedom. 

These degrees of freedom are, in principle, unbounded, while any practical detector for measuring them only has a finite range of detection, so a natural question has been whether the potential for the state to fall beyond the range of detection poses any serious consequences for the security of a protocol \cite{Qi2011,Nunn2013,Ray2013a,Ray2013,Toscano2018,Qin2015,Qin2018}. Qi first noted the potential for a detection range loophole in time-frequency QKD \cite{Qi2011}, with Nunn \textit{et. al.} outlining a specific strategy for exploiting the loophole \cite{Nunn2013}. In the context of both time-frequency QKD and more general continuous variable entanglement verification and quantum cryptographic protocols, Ray and van Enk discussed how data falling outside the measurement range compromises variance-based measures of entanglement, and demonstrated that R\'enyi entropies without quantum memory provide more optimistic bounds for verifying entanglement \cite{Ray2013a,Ray2013}. The issues arising from finite measurement ranges have additionally been discussed in \cite{Furrer2012,Qin2015,Qin2018}, in the context of homodyne-based continuous variable (CV) QKD\footnote{CV QKD commonly refers to protocols employing electric field quadratures.}, with the main problem being caused by detections above the saturation limit of the detector. A major review of CV QKD is available in Section VI of \cite{Weedbrook2012}. 

In Section 2, we review the entropic uncertainty relations with quantum memory, as well as how results outside the measurement range can render the relationship trivial. In Section 3, we consider a more general problem: given a measurement outcome one would like to safely ignore, such as the state falling outside the measurement range, we formulate a non-trivial entropic uncertainty relation with quantum memory that depends on the probability of the problematic outcome, rather than on the operators characterizing the problematic outcome. This bound is particularly important for entanglement-based high-dimensional quantum cryptography protocols. In Section 4, we discuss the modified bound in the context of time-frequency QKD, and find that additional assumptions are required to deal with practical limitations like loss and vacuum components. Even with some additional assumptions about the source and our modified result, channel loss severely limits the secure key rate of the protocol. Finally, in Section 5, we discuss the applicability of our main result to entanglement-based CV QKD using homodyne detection. We find that, since being outside the measurement range corresponds to saturation, not loss, our bound produces the same results as existing protocols with the added benefit that it provides a quantitative way to guard against saturation attacks.

\section{Entropic uncertainty relations and the measurement range problem}

We begin this section by summarizing the unmodified entropic uncertainty relations, (a thorough review of entropic uncertainty relations and their applications can be found in \cite{Coles2017}). We consider a source, potentially controlled by Eve, that distributes quantum systems $A$ and $B$ to Alice and Bob, respectively, so that the total purified state is $\rho_{ABE}$. Alice either performs a Z-type measurement, characterized by the positive operator-valued measure (POVM), $Z=\{\mathbb{Z}_{A}^{z}\}_{z}$, or an X-type measurement, characterized by the POVM, $X=\{\mathbb{X}_{A}^{x}\}_{x}$. Bob also alternates between the same two types of measurements. See Figure 1 for an illustration of the set-up.

\begin{figure}
\centering
\includegraphics[scale=0.27]{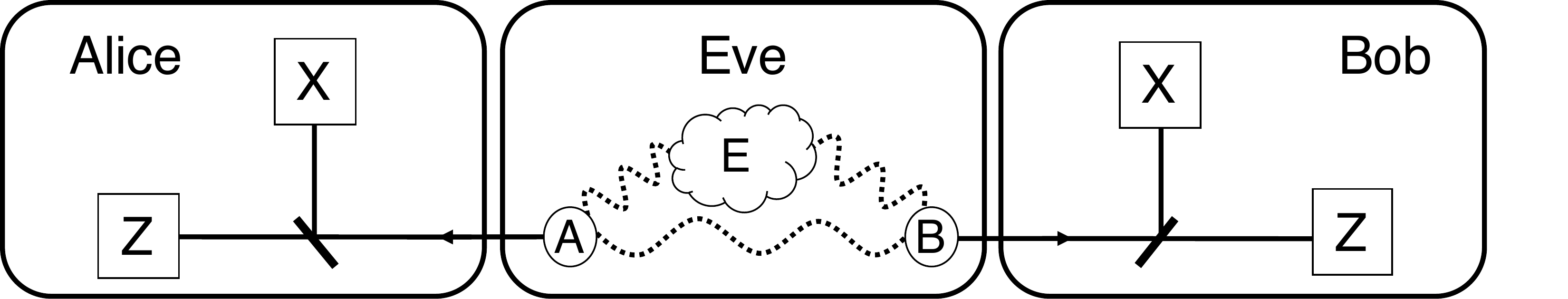}

\caption{A typical scenario for the entropic uncertainty relations with quantum memory. Alice and Bob are given quantum systems, $A$ and $B$, from a source controlled by Eve, who has quantum memory, $E$. Alice and Bob independently and randomly either make X-type or Z-type measurements.}

\end{figure}

The most information Eve can have about Alice's Z-type measurement results is quantified by the conditional min-entropy of Alice's classical register, $Z_{A}$, used for storing results from Z-type measurements, given Eve's quantum memory, $E$:
\begin{equation}
H_{\text{min}}(Z_{A}|E)=-\log p_{\text{guess}}(Z_{A}|E)
\end{equation}
where $p_{\text{guess}}(Z_{A}|E)=\max_{\mathbb{M}_{E}}\sum_{z}p_{Z_{A}}^{z}\text{Tr}(\mathbb{M}_{E}^{z}\rho_{E}^{z})$
is Eve's guessing probability of $Z_{A}$ maximized over all POVMs
on $E$, $\mathbb{M}_{E}$ \cite{Renner2005,Konig2009,Coles2017}.

The entropic uncertainty relations with quantum memory provide a bound on $H_{\text{min}}(Z_{A}|E)$
\cite{Berta2010,Tomamichel2011,Coles2012}:
\begin{equation}
H_{\text{min}}(Z_{A}|E)+H_{\text{max}}(X_{A}|B)\geq-\log c(X,Z)
\end{equation}
where $H_{\text{max}}(X_{A}|B)$ is the conditional max-entropy of Alice's classical register, $X_{A}$, for storing results from X-type measurements, given Bob's system, $B$. It is given by \cite{Coles2012}:
\begin{equation}
H_{\text{max}}(X_{A}|B)=2\log\max_{\sigma_{B}}F(\rho_{X_{A}B},\mathbb{I}_{X_{A}}\otimes\sigma_{B})
\end{equation}
with $F(\rho,\sigma)=Tr(\sqrt{\sqrt{\sigma}\rho\sqrt{\sigma}})$ denoting the fidelity between operators $\rho$ and $\sigma$. $H_{\text{max}}(X_{A}|B)$ can be upper bounded using $H_{\text{max}}(X_{A}|X_{B})$, in which the max-entropy is conditional on Bob's results from an X-type measurement \cite{Frank2013}. Finally, the bound depends on the POVM elements for the two measurements. One bound is the maximum overlap between the two POVMs:
\begin{equation}
c(X,Z)=\max_{x,z}||\sqrt{\mathbb{X}_{A}^{x}}\sqrt{\mathbb{Z}_{A}^{z}}||_{\infty}^{2}
\end{equation}
where $||\cdot||_{\infty}$ denotes the maximum singular value \cite{Berta2010,Coles2012}.
An equal or better bound is provided by \cite{Tomamichel2012a}:
\begin{equation}
c'(X,Z)=\min\big\{\max_{x}||\sum_{z}\mathbb{Z}_{A}^{z}\mathbb{X}_{A}^{x}\mathbb{Z}_{A}^{z}||_{\infty},\max_{z}||\sum_{x}\mathbb{X}_{A}^{x}\mathbb{Z}_{A}^{z}\mathbb{X}_{A}^{x}||_{\infty}\big\}.
\end{equation}

Given the reliance of the bound on the POVMs, a problem arises when Alice has POVM elements from $X$ and $Z$ that cause $c(X,Z)\approx 1$. The general problem can be summarized as follows: Alice has two POVMs, $Z=\{\mathbb{Z}_{A}^{z}\}_{z=1}^{N_{Z}}\cup\{\mathbb{Z}_{A}^{\emptyset}\}$ and $X=\{\mathbb{X}_{A}^{x}\}_{x=1}^{N_{X}}\cup\{\mathbb{X}_{A}^{\emptyset}\}$, such that $||\sqrt{\mathbb{Z}_{A}^{\emptyset}}\sqrt{\mathbb{X}_{A}^{\emptyset}}||_{\infty}^{2}\approx 1$, while $||\sqrt{\mathbb{Z}_{A}^{z}}\sqrt{\mathbb{X}_{A}^{x}}||_{\infty}^{2}<1$ for the other POVM elements. The elements denoted with ``$\emptyset$'' indicate some measurement outcomes that cause $c(X,Z)\approx 1$. For convenience, we will often refer to these as ``null'' measurement outcomes, since a common way for $c(X,Z)\approx 1$  is when the detector does not register a result; however, we do not specify a form for these elements, meaning in general they can correspond to any measurement outcomes with POVM elements that are problematic for the maximum overlap. The current entropic uncertainty relation would  provide a trivial bound due to these null measurements, but ideally we would like to salvage a bound using the other POVM elements that do not saturate the overlap. In the next section we provide a modified entropic uncertainty relation in which Alice and Bob can still bound Eve's information about the Z-type measurement outcomes, in terms of the X-type measurement outcomes for $1\leq x\leq N_{X}$, the POVM elements from $\{\mathbb{Z}_{A}^{z}\}_{z=1}^{N_{Z}}$ and $\{\mathbb{X}_{A}^{x}\}_{x=1}^{N_{X}}$, and the probability Alice's measurements return null outcomes.

Before introducing the bound, we illustrate the relevance of this limitation to current entropic uncertainty relations by considering what occurs when Alice has basis-dependent limitations on her measurement range \cite{Qi2011,Furrer2012,Nunn2013,Ray2013a,Ray2013}. For instance, if Alice were measuring the frequency of a single photon, she might have $N_{Z}$ POVM elements for describing the $N_{Z}$ frequency bins that the detector can resolve, plus the null measurement POVM element, $\mathbb{Z}_{A}^{\emptyset}$, to characterize the result of the photon falling outside the bandwidth of the detector. If her X-type measurement corresponds to an arrival-time measurement, then she will also have a finite number of bins with good timing resolution, then a null measurement element, $\mathbb{X}_{A}^{\emptyset}$, corresponding to the case when the photon arrived before or after her well-resolved time bins. One might think that because frequency and arrival-time are non-commuting observables, Alice and Bob would be able to have a non-trivial bound on Eve's information. Unfortunately,
in such a case, because the null measurement POVM elements are measurement-dependent (i.e. $\mathbb{Z}_{A}^{\emptyset}\neq\mathbb{X}_{A}^{\emptyset}$) and span a semi-infinite region of the Hilbert space, $c(X,Z)=||\sqrt{\mathbb{Z}_{A}^{\emptyset}}\sqrt{\mathbb{X}_{A}^{\emptyset}}||_{\infty}^{2}\approx1$.

To resolve this issue, one might be tempted to discard all the null outcomes from the frequency and arrival-time measurements, and bound the conditional min-entropy of the remaining data from the frequency measurement using the remaining data from the arrival-time measurement. By only maximizing the overlap over the POVM elements corresponding
to the well-resolved bins, perhaps one can discount the null measurement POVM elements. However, a conceptually simple strategy, outlined in \cite{Nunn2013}, can be used by Eve to make Alice and Bob overestimate the conditional min-entropy: Eve can make measurements of the photon frequency with very narrow bin widths, such that when Alice and Bob perform arrival-time
measurements, the most likely outcome is a null measurement. Thus, Eve gains all information about frequency without introducing any errors to their frequency measurements, and Alice and Bob will almost always discard the arrival-time events for which Eve does not have any information
and which would otherwise be used to bound Eve's information.

Note that this problem occurs when the null measurement POVM element depends on the measurement type, i.e. when $\mathbb{X}_{A}^{\emptyset}\neq\mathbb{Z}_{A}^{\emptyset}$ \cite{Tomamichel2012}. In Appendix 1, we show that, when the null measurement POVM element is the same in both measurements, which can naively lead to $c(X,Z)=1$, one can always formulate an entropic uncertainty relation in terms of the POVM elements of a related, effective measurement that does not have the problematic null measurement element. We show how to construct the related, effective measurement, and discuss why the approach for solving the problem when the null measurement is basis-independent fails when the null measurement is basis-dependent.

\section{A modified entropic uncertainty relation}
Our main result is a modified entropic uncertainty relation for a scenario in which Alice has specific outcomes in Z- and X-type measurements for which the POVM elements characterizing those outcomes yield a trivial bound on Eve's information, but those outcomes have low probability of occuring. Our result allows Alice and Bob to achieve a sometimes better bound on Eve's information about the measurement outcomes, with no extra characterization of the state required, simply by including the fact that the problematic outcomes have a low probability of occuring.

\emph{Main result }: For a tripartite state, $\rho_{ABE}$, and two POVMs on $\mathcal{H}_{A}$, $Z=\{\mathbb{Z}_{A}^{z}\}_{z=1}^{N_{Z}}\cup\{\mathbb{Z}_{A}^{\emptyset}\}$ and $X=\{\mathbb{X}_{A}^{x}\}_{x=1}^{N_{X}}\cup\{\mathbb{X}_{A}^{\emptyset}\}$,
\begin{equation}
H_{\text{min}}(Z_{A}|E)\geq-2\log\left[\sqrt{p_{Z_{A}}^{\emptyset}}+\sqrt{p_{X_{A}}^{\emptyset}}+\sqrt{1-p_{X_{A}}^{\emptyset}}\sqrt{c^{<}(X,Z)}\left(\sqrt{2}^{H_{\text{max}}(X_{A}^{<}|B)}\right)\right]
\end{equation}
where $p_{Z_{A}}^{\emptyset}=\text{Tr}(\rho_{A}\mathbb{Z}_{A}^{\emptyset})$ and $p_{X_{A}}^{\emptyset}=\text{Tr}(\rho_{A}\mathbb{X}_{A}^{\emptyset})$ are Alice's probabilities of null measurements from Z- and X-type measurements. $H_{\text{max}}(X_{A}^{<}|B)$ is the conditional max-entropy of Alice's X-type measurement results, after she has discarded the null measurements, given Bob's system. Finally,
\begin{equation}
c^{<}(X,Z)=\max_{(x,z)\neq\emptyset}||\sqrt{\mathbb{X}_{A}^{x}}\sqrt{\mathbb{Z}_{A}^{z}}||_{\infty}^{2}
\end{equation}
is the maximum overlap for the Z and X POVM elements, excluding the null measurement POVM elements that cause the original bound to saturate.

Practically, the quantities in Eq. (6) are as experimentally accessible as those in Eq. (2), the unmodified entropic uncertainty relation. The new maximum overlap in Eq. (7) can be calculated after having characterized the POVMs for the detectors \cite{Enk2017}. The null measurement probabilities can be calculated directly from the statistics of the measurement outcomes. $H_{\text{max}}(X_A^<|B)$ can be bounded from above by $H_{\text{max}}(X_A^<|X_B)$, the conditional max-entropy of Alice's X type measurement, after she has discarded the null measurements, given Bob's X measurement results. As we discuss in Section 4, Bob may need to make some modifications to his data, like replacing his null outcomes with bit values fitted to Alice's distribution. $H_{\text{max}}(X_A^<|X_B)$ can be further bounded from above using methods from \cite{Niu2016}.

The result is quite general: we make no assumptions about the dimension of the Hilbert spaces or the commutation relations between the POVM elements. The proof of this result is available in Appendix 2. We use similar analytical techniques from \cite{Coles2012}, which provides a proof of Eq. (2). Additionally, we use the form of the fidelity function in terms of trace norm, $F(\rho,\sigma)=||\sqrt{\rho}\sqrt{\sigma}||_{\text{Tr}}$, and exploit the triangle inequality property for norms. In Appendix 3, we provide a smoothed version of the bound, which could be used to include finite-size effects, as Eq. (6) is valid for the asymptotic limit.

As stated above, our result provides a \textit{sometimes} better bound on $H_{\text{min}}(Z_{A}|E)$, and so a few comments are in order on the applicability of the result. First, note that if $\sqrt{p_{Z_{A}}^{\emptyset}}+\sqrt{p_{X_{A}}^{\emptyset}}$ exceeds 1 then the trivial lower bound of $H_{\text{min}}(Z_A|E)\geq 0$ is better. See Figure 2 for a contour plot of necessary values of $p_{Z_{A}}^{\emptyset}$ and $p_{X_{A}}^{\emptyset}$ for a non-trivial bound. Having reasonable values of $p_{X_{A}}^{\emptyset}$ and $p_{Z_{A}}^{\emptyset}$ depends on what $\mathbb{X}_{A}^{\emptyset}$ and $\mathbb{Z}_{A}^{\emptyset}$ correspond to physically. For example, in Section 4, in the context of time-frequency QKD, these null measurement probabilities should ideally correspond to the probability that single photons arriving at Alice's detectors fall outside her measurement ranges, resulting in no detector clicks; however, in practice, there are many scenarios that also result in null detection, including loss and vacuum components. While these other scenarios do not compromise security, they certainly cause an overestimation of $p_{X_{A}}^{\emptyset}$ and $p_{Z_{A}}^{\emptyset}$ and a pessimistic bound on the secure key rate. We will discuss what \textit{additional} assumptions may need to be made to achieve a good bound. In Section 5, when we discuss homodyne-based CV QKD, $\mathbb{X}_{A}^{\emptyset}$ and $\mathbb{Z}_{A}^{\emptyset}$  correspond to quadrature intensities above the saturation limit of the detectors; in practice, saturation probabilities can be kept low, so fewer assumptions are needed to achieve a good bound.

\begin{figure}
\centering
\includegraphics[scale=0.048]{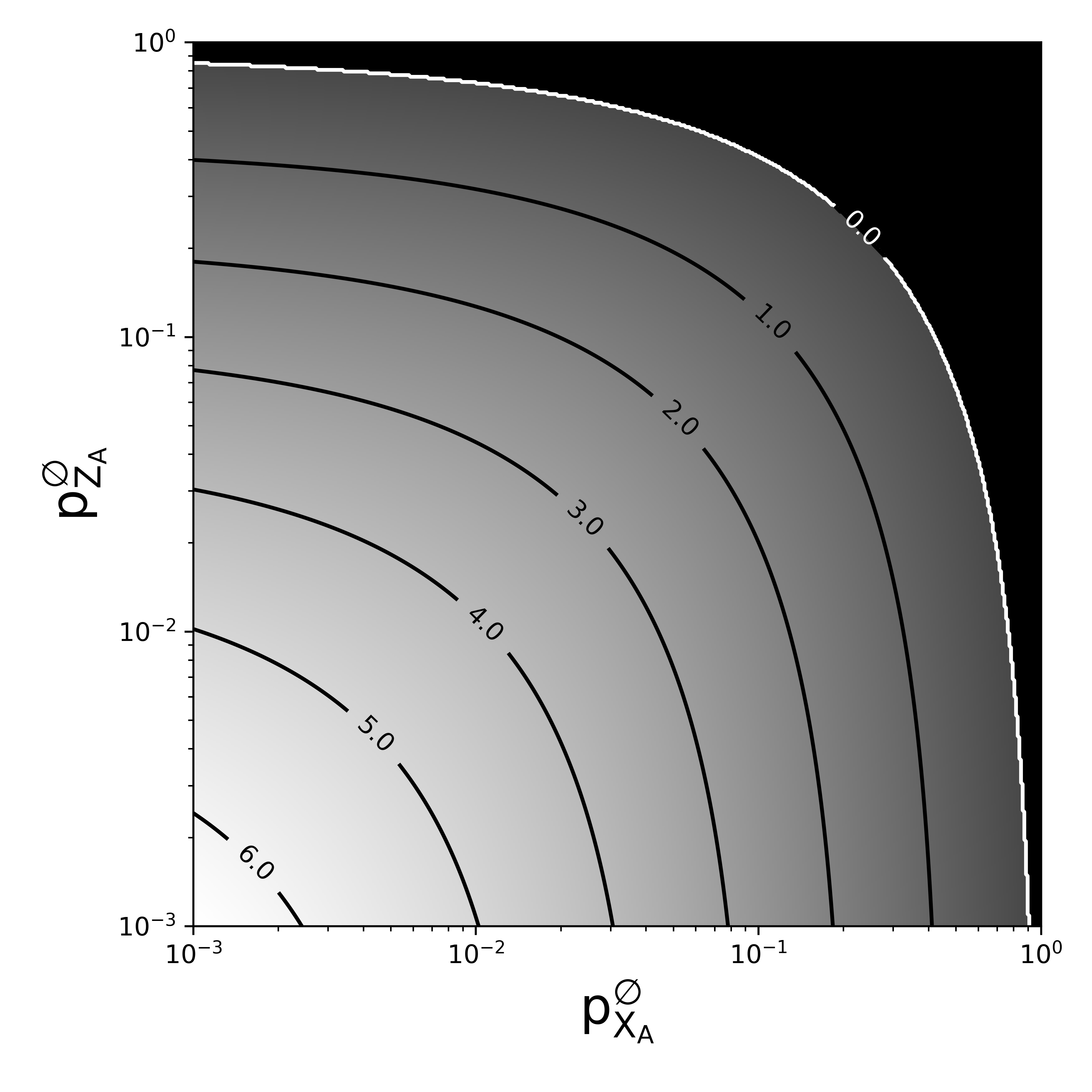}
\caption{Contour plot of the lower bound, in bits, on $H_{\text{min}}(Z_{A}|E)$ as a function of $p_{Z_{A}}^{\emptyset}$ and $p_{X_{A}}^{\emptyset}$ as given by Eq. (6). We have fixed $H_{\text{max}}(X_{A}^{<}|B)=1$, which corresponds to a noiseless scenario, and $c^{<}=10^{-3}$, which is an experimentally feasible value in time-frequency QKD \cite{Niu2016}. This plot corresponds to best-case scenarios for the bound, and provides necessary conditions on the values of $p_{Z_{A}}^{\emptyset}$ and $p_{X_{A}}^{\emptyset}$ for protocols employing Eq. (6) to prove security, with the black region corresponding to no proven security. Assuming no noise, for $c^{<}=10^{-3}$, the maximum tolerable equal probability of null measurement is $p_{Z_{A}}^{\emptyset}=p_{X_{A}}^{\emptyset}\approx23.2\%$, while in the limit $c^{<}\rightarrow0$, $p_{Z_{A}}^{\emptyset}=p_{X_{A}}^{\emptyset}=25\%$. Of course, if we make $p_{Z_{A}}^{\emptyset}$ smaller, we gain tolerance for higher $p_{X_{A}}^{\emptyset}$, and vice versa. For example, with $p_{Z_{A}}^{\emptyset}=10^{-3}$, $p_{X_{A}}^{\emptyset}$ can be as high as $\sim$92\%.}

\end{figure}

Second, to tolerate higher null measurement probabilities, one requires low values of $c^{<}(X,Z)$ and $H_{\text{max}}(X_{A}^{<}|B)$. The former can be achieved if the degree of freedom being measured is high-dimensional and the Z- and X-type measurements are non-commuting, while the latter can be achieved if Alice and Bob's systems are highly entangled. We will show in the next two sections that our bound provides insight into the security proofs for entanglement-based time-frequency, and continuous variable QKD using homodyne detection, two protocols employing high-dimensional degrees of freedom and Fourier pair measurements.

\section{Application to time-frequency QKD}

One of the main applications for entropic uncertainty relations is in proofs of the security of QKD protocols. QKD is a method for two spatially separated parties, Alice and Bob, to establish a shared, secret cryptographic key in the presence of an eavesdropper, Eve, where, for instance, the results of one measurement can be used to form the key, and the results of the other measurement can be used to bound Eve's information about the key \cite{Scarani2009}. A distinction is made between prepare-and-measure protocols, in which Alice uses a fully-characterized source to send signals to Bob, who has fully-characterized measurements, and entanglement-based protocols, in which Alice and Bob both have full characterization of their measurements, and an untrusted source outputs states for them to measure. The prepare-and-measure scenario can usually be framed in terms of the entanglement-based scenario \cite{Bennett1992}, and the latter is suited to be analyzed using entropic uncertainty relations, since the statistics of the measurement outcomes and the POVMs characterizing the measurement can be used to bound Eve's information.

One example of a QKD protocol for which our main result is particularly relevant is time-frequency QKD. Time-frequency QKD employs single photon frequency and arrival-time as non-commuting observables to establish a key \cite{Qi2006}. In an entanglement-based protocol, an untrusted source would output two spatial modes, one sent to Alice and one to Bob, who would then randomly alternate between measuring frequency and arrival-time, with each observable having bins of finite width into which results can fall. 

Typically, Alice's frequency POVM in the single photon subspace, $F_{\text{ideal}}$, is idealized with elements \cite{Nunn2013}:
\begin{equation}
\mathbb{F}_{A}^{m}=\int_{\omega_{m}-\delta\omega/2}^{\omega_{m}+\delta\omega/2}\frac{d\omega}{2\pi}|\omega\rangle\langle\omega|
\end{equation}
where $\omega_{m}$ is the central frequency of the bin; $\delta\omega$ is the bin width; and $|\omega\rangle$ is a frequency eigenstate with normalization $\langle\omega|\omega'\rangle=2\pi\delta(\omega-\omega')$ \cite{Niu2016}. It is often assumed that the central frequencies of the bins are defined over positive and negative frequencies \cite{Nunn2013,Zhang2014}, although some have modified the definition to include only positive frequencies \cite{Niu2016}. Of course, a full characterization of Alice's measurement will also include POVM elements for vacuum and multiphoton contributions; we will return to this issue later in our discussion. 

The arrival-time eigenstates are defined as Fourier pairs of the frequency eigenstates, with Alice's arrival-time POVM in the single photon subspace, $T_{\text{ideal}}$, typically defined by elements \cite{Niu2016}:
\begin{equation}
\mathbb{T}_{A}^{k}=\int_{t_{k}-\delta t/2}^{t_{k}+\delta t/2}dt|t\rangle\langle t|,\ 
\end{equation}
where $t_{k}$ is the central arrival time of the bin; $\delta t$ is the bin width; and $|t\rangle=\int_{-\infty}^{+\infty}\frac{d\omega}{2\pi}e^{i\omega t}|\omega\rangle$ is the Fourier transform of $|\omega\rangle$. As with frequency, the sequence of central times extends over positive and negative arrival times. 

Using Eq. (4), these POVMs yield a maximum overlap of:
\begin{equation}
c(T_{\text{ideal}},F_{\text{ideal}})=\frac{\delta\omega\delta t}{2\pi}S_{0}^{(1)}(1,\frac{\delta\omega\delta t}{4})^{2}
\end{equation}
where $S_{0}^{(1)}(\cdot,\cdot)$ is the 0th radial prolate spheroidal wave function of the first kind \cite{Rudnicki2012,Coles2017}. Eq. (10) is roughly proportional to $\delta\omega\delta t$ when their product is small \cite{Rudnicki2012}. As expected, $c(T_{\text{ideal}},F_{\text{ideal}})\rightarrow1$ as the bin width product grows.

The obvious problem with this characterization is that realistic detectors cannot have a constant bin width over all frequencies and arrival-times. For instance, a typical source for time-frequency entangled photons may have a repetition rate on the order of 10 MHz \cite{Lee2014,Chen2017}, which would limit the range of arrival-times that can be employed. Additionally, the best single photon detectors have high efficiency in a finite spectral range of a few tens of nanometers about 1550 nm \cite{Marsili2013}. We assume that when the photon falls beyond these ranges of the detectors, the detectors do not click. Thus, one step towards a more realistic characterization of the measurements is to include the null measurement elements corresponding to when the photon falls outside the temporal range in the arrival-time measurement or the spectral range in the frequency measurement. If the temporal range is from $[-t_{c},t_{c}]$, then $\mathbb{T}_{A}^{\emptyset}=\int_{-\infty}^{-t_{c}}dt|t\rangle\langle t|+\int_{t_{c}}^{+\infty}dt|t\rangle\langle t|$. If the spectral range is from $[\omega_{o}-\omega_{c},\omega_{o}+\omega_{c}]$, then $\mathbb{F}_{A}^{\emptyset}=\int_{-\infty}^{\omega_{o}-\omega_{c}}\frac{d\omega}{2\pi}|\omega\rangle\langle\omega|+\int_{\omega_{o}+\omega_{c}}^{+\infty}\frac{d\omega}{2\pi}|\omega\rangle\langle\omega|$. Thus, the refined POVMs are $T=\{\mathbb{T}_{A}^{k}\}_{k=1}^{N_{T}}\cup\{\mathbb{T}_{A}^{\emptyset}\}$ and $F=\{\mathbb{F}_{A}^{m}\}_{m=1}^{N_{F}}\cup\{\mathbb{F}_{A}^{\emptyset}\}$, where we assume that $t_{c}$ ($\omega_{c}$) or $\delta t$ ($\delta\omega$) can be chosen so that the $N_{T}$ ($N_{F}$) bins can be defined within $[-t_{c},t_{c}]$ ($[\omega_{o}-\omega_{c},\omega_{o}+\omega_{c}]$). Based on (10), $c(T,F)\approx1$ since the null measurement bins have, in principle, infinite width. As discussed previously, this has been a known issue for time-frequency QKD \cite{Qi2011,Nunn2013}.

A common suggestion for dealing with this problem has been to apply pre-measurement filters to exclude the frequencies or arrival-times that would not be detected \cite{Qi2011,Nunn2013}. Consider, however, that for the unmodified bound, the maximum overlap in Eq. (4) is not sensitive to the probabilities of different measurement outcomes, only to the POVM elements. This means that, as Eq. (2) stands, it does not necessarily matter that a filter could keep those probabilities low. Thus, without modifying the entropic uncertainty relations, the only way to ensure that we can safely disregard the null measurement POVM elements with semi-infinite support that saturate Eq. (4) is to ensure that the state has no shared support with those POVM elements, effectively reducing the problem to the smaller subspace defined by the support of the state. Unfortunately, it is not possible to construct a filter that could simultaneously ensure  compact support in the time domain and frequency domain \cite{Rudin1987}. In other words, we cannot simultaneously make the state have exactly zero probability of yielding null measurements on the spaces of $\mathbb{T}_{A}^{\emptyset}$ and $\mathbb{F}_{A}^{\emptyset}$ without filtering out all of the state. As long as there is some non-zero probability, one would need to consider the POVM elements on that space, which saturate Eq. (4) and render the unmodified entropic uncertainty relations trivial.

Alternatively, as discussed in \cite{Ray2013,Ray2013a}, one could assume that there exists a cut-off outside the measurement range, and that the probability of a result beyond the cut-off is negligible, using less coarse-grained measurements with a wider range to estimate this probability. We would then need to ask how small that probability needs to be, and how to quantify its effect within the bound on security.

Our main result, Eq. (6), is a step towards dealing with this problem. In \cite{Niu2016}, an ideal source for time-frequency entangled photons has a biphoton wavefunction modelled as:
\begin{equation}
\psi(\omega_{A},\omega_{B})=\frac{\exp[-(\omega_{A}-\omega_{B})^{2}\sigma_{cor}^{2}/4-(\omega_{A}+\omega_{B})^{2}\sigma_{coh}^{2}]}{\sqrt{\pi/2\sigma_{coh}\sigma_{cor}}}
\end{equation}
where $\omega_{A}$ and $\omega_{B}$ are relative to the central telecom frequency; $\sigma_{coh}$, the coherence time, is taken to be 6 ns, and $\sigma_{cor}$, the correlation time, is taken to be 2 ps. Assuming the spectral range is between 1520 nm and 1610 nm \cite{Marsili2013}, the state in Eq. (11) yields a probability for a null outcome in the frequency measurement of $p_{F_{A}}^{\emptyset}=0$ (to within machine precision\footnote{We used MATLAB, for which a positive value less than $2^{-52}$ is treated as 0 \cite{Moler1996}.}). Note that this is in the best case scenario when an eavesdropper is not tampering with the value; of course, in a real-world entanglement-based QKD scenario this probability value would be measured from the data. Using a repitition rate of 55.6 MHz \cite{Niu2016}, Eq. (11) yields a probability for a null outcome in the arrival-time measurement of $p_{T_{A}}^{\emptyset}\approx0.27\%$. We can upper bound $c^{<}(T,F)=\max_{k,m\neq\emptyset}||\sqrt{\mathbb{T}_{A}^{k}}\sqrt{\mathbb{F}_{A}^{m}}||_{\infty}^{2}$ by $c(T_{\text{ideal}},F_{\text{ideal}})$ since the sets of POVMs over which the former is maximized are subsets of the sets over which the latter is maximized. Again using values from \cite{Niu2016}, this yields $c^{<}\leq10^{-3}$.

The original problem was that the unmodified entropic uncertainty relation, Eq. (2), was rendered trivial by the overlap $||\sqrt{\mathbb{Z}_{A}^{\emptyset}}\sqrt{\mathbb{X}_{A}^{\emptyset}}||_{\infty}^{2}\approx1$, so it is a clear improvement that the new maximum overlap, $c^{<}$, is no longer saturated by the null measurement POVM elements. Unfortunately, a new problem arises when applying Eq. (6) in a practical implementation of entanglement-based, time-frequency QKD: $p_{F_{A}}^{\emptyset}$, $p_{T_{A}}^{\emptyset}$, and $H_{\text{max}}(X_{A}^{<}|B)$ will all need to be estimated using the measurement data, a task limited by source, coupling, and channel imperfections.

In a true entanglement-based scenario, only the detectors are trusted. This means that the source, channels and couplings are not. In Eq. (6), $p_{F_{A}}^{\emptyset}$ and $p_{T_{A}}^{\emptyset}$ should ideally correspond to the probabilities that the single photon portion of Alice's state yields null measurements. However, less than ideal devices may result in an overestimation of these parameters. 

First, the source will pose a problem due to vacuum components since since vacuum states also yield null measurements. For example, a common choice for creating spectrally entangled photon pairs is a spontaneous parametric down-conversion device, which can have low conversion efficiency \cite{Zhong2015}. Therefore, to get a better estimate on $p_{F_{A}}^{\emptyset}$ and $p_{T_{A}}^{\emptyset}$, Alice would ideally want to determine the probability the untrusted source emits a vacuum state. While passive decoy state methods allow for the characterization of the photon number distribution for untrusted sources in single-mode QKD \cite{Zhao2010}, we are not aware of any methods for characterizing the photon number distribution of an untrusted source for high-dimensional QKD employing many time-frequency modes. Thus, short of such methods, the source will be untrusted and uncharacterized, and we would have to allow null measurements due to vacuum components to be part of the estimation of $p_{F_{A}}^{\emptyset}$ and $p_{T_{A}}^{\emptyset}$. However, as shown in Figure 2, this would mean the probability of a vacuum component would need to be less than 25\%, and that is additionally on condition that there be perfect correlations in measurement results, and a very low value of $c^{<}$. This may not be possible with current technology, as increasing the number of photon pairs is typically achieved by pumping at higher intensities which also results in increasing the number of multiphoton contributions \cite{Zhong2012}. Multiphoton detection events are a source of noise \cite{PhysRevA.61.052304}, and thus the high correlations between Alice and Bob's data will be unattainable. 

Second, the problem caused by in-lab coupling from the channels to the detectors is conceptually similar: a lost photon yields a null measurement. Even if the couplings induce a basis-independent loss, if we do not want to trust the couplings, we will need to factor their contribution to null measurements into the estimation of $p_{F_{A}}^{\emptyset}$ and $p_{T_{A}}^{\emptyset}$. To have a non-trivial bound, that loss cannot be greater than 25\%, on condition that vacuum contributions are negligible, correlations between measurement results are high, and we have a very low value of $c^{<}$. 

There is clearly a practicality problem if we do not trust the source since there are so many scenarios that can cause a null measurement. The problem can be slightly improved if we allow partial characterization of the source and in-lab couplings, weakening the full entanglement-based assumption. For instance, Alice need not know the full biphoton wavefunction, but if she can know the probability that the source outputs a vacuum state and characterize the basis-independent coupling losses between the source and her detector, then she can have a much better estimate of $p_{F_{A}}^{\emptyset}$ and $p_{T_{A}}^{\emptyset}$. This is because she can first use the process from Appendix 1 to discount any negative impact due to basis-independent null measurement POVM elements, and then she can use Eq. (6) to treat the remaining, basis-dependent null measurements, whose impact on security is treated via the probabilities $p_{F_{A}}^{\emptyset}$ and $p_{T_{A}}^{\emptyset}$. With no tampering of her single photon component, she should expect the very low values for $p_{F_{A}}^{\emptyset}$ and $p_{T_{A}}^{\emptyset}$ calculated earlier. 

However, in all QKD protocols the channel is untrusted. In Eq. (6), $H_{\text{max}}(X_{A}^{<}|B)$ should ideally be the conditional max-entropy of Alice's measurement outcomes from single photon components that fell within her measurement range, given the single photons that arrived at Bob's detector. Unfortunately, Bob will have additional null measurements due to all the photons lost in the channel. Even if Bob can discard null measurements due to the source outputting a vacuum component, or due to coupling from the channel to his detector, he will have to assume all the null measurements due to loss in the channel are instead due to single photons falling outside his detector range. Since there will be cases where Alice did not have a null measurement while Bob did, Bob's strategy in those cases will then be to assign a bit value to these null outcomes based off the publicly known probability distribution of Alice's results, to have a lower chance of error.

Using the state in Eq. (11), with a loss of 0.2 dB/km in fiber, assuming Alice has partial characterization of her source so she can safely estimate her null measurement probabilities to be $p_{F_{A}}^{\emptyset}=0$ and $p_{T_{A}}^{\emptyset}=0.27\%$, and taking $c^{<}=10^{-3}$, the bound on $H_{\text{min}}(F_{A}|E)-H(F_{A}|F_{B})$ still saturates at $\sim$2 km, even in an ideal case where Alice and Bob have no dark counts and Eve has not interfered with the results, as shown in Figure 3. Here, $H(F_{A}|F_{B})$ is the conditional Shannon entropy of Alice's frequency results given Bob's, and it is used to quantify the number of bits to correct errors in the key \cite{Scarani2009,Nunn2013}; in this case, it will be non-zero even in noiseless channels because of the finite coherence time of the state. We used methods from \cite{Niu2016} to bound $H_{\text{max}}(T_{A}^{<}|B)$. Note that our result significantly differs from the distance of >150 km presented in \cite{Niu2016}; however, their analysis did not address the measurement range problem, so it would not provide security given realistic limitations on the measurement range. We have shown the region of security can be expanded, albeit slightly, to more than 2 km.

\begin{figure}
\centering
\includegraphics[scale=0.06]{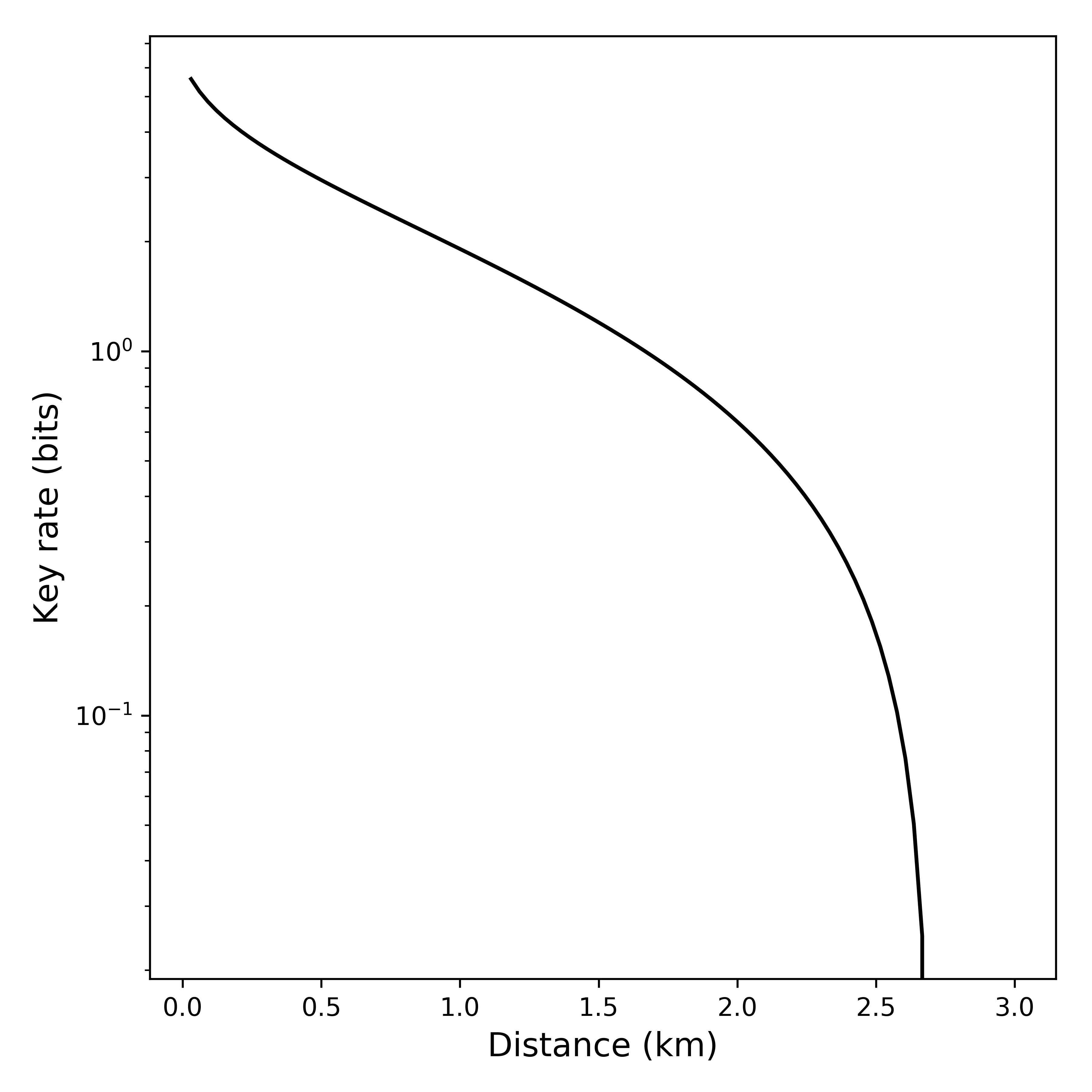}

\caption{Key rate vs. distance for time-frequency QKD, accounting for security repercussions of null measurements on Bob's side. For the state in Eq. (11), we have set $\sigma_{coh}=6\mathrm{ns}$, and $\sigma_{cor}=2\mathrm{ps}$ \cite{Niu2016}. To employ Eq. (6) we have used $c^{<}=10^{-3}$, $p_{T_{A}}^{\emptyset}=0.27\%$, and $p_{F_{A}}^{\emptyset}=0\%$. We have bounded $H_{\text{max}}(T_{A}^{<}|B)$ using techniques from \cite{Niu2016}, employing a bin width of $\delta t=20\mathrm{ps}$. See the text for additional assumptions. Previous security analysis of time-frequency QKD in \cite{Niu2016} presented non-zero key rate at distances of >150 km, but the proof does not provide security against the measurement range problem. Our new bound widens the region of security to more than 2 km.}

\end{figure}

It appears that bounding the security of entanglement-based time-frequency QKD with a completely untrusted source is still impractical using Eq. (6). With knowledge of the source's vacuum component probability, and characterization of their coupling losses, Alice and Bob can achieve a better bound on $H_{\text{min}}(F_{A}|E)$, but only if the channel distance to Bob is very short. The result may still be useful for applications of time-frequency entanglement other than entanglement-based communication, such as entanglement witnessing \cite{Berta2010}.

Clearly, time-frequency QKD with arrival-time and frequency treated as Fourier pairs of each other is not feasible with an untrusted source and current levels of loss; seeing that the null measurements depend on the type of measurement, loss must be treated as a threat to security. We therefore have the following outlook for time-frequency QKD: if one wants to continue characterizing arrival-time and frequency as Fourier pairs, then one will need to move to a prepare-and-measure or a measurement-device-independent QKD setting. For the former approach, the equivalence between prepare-and-measure and entanglement-based QKD that is often used for security proofs will need to be carefully considered, since an entanglement-based approach is untenable. Alternatively, it may be fruitful to re-examine the characterization of arrival-time and frequency measurements: it is likely that one is not measuring perfect Fourier pairs in the lab, so perhaps the newly-characterized observables that one is measuring will end up having basis-independent null measurements, which we know not to be a problem for security. Finally, some work has been done to implement measurements in arrival-time-like and frequency-like measurements: rather than measuring Fourier pairs of observables, one either makes a measurement in a basis of $d$ arrival times, $\{|t_n\rangle\}_{n=0}^{d-1}$, or in a superposition basis, $\{|f_{m}\rangle=\sum_{n=0}^{d-1}\exp(2\pi inm/d)|t_n\rangle\}_{m=0}^{d-1}$ \cite{Islam2017,Islam2017a}. In that case, the measurement range problem is averted because the system of interest is now defined on an effective subspace in which there are lower probabilities of basis-dependent null measurements.

\section{Application to homodyne-based continuous variable QKD}

CV QKD exploits the electric field quadratures as the non-commuting
observables to establish a secret key. In a typical entanglement-based protocol employing homodyne detection, an untrusted source sends one mode to Alice, another to Bob, and they each randomly perform either $X$ or $P$ quadrature measurements \cite{Grosshans2003,Laudenbach2018}. Entanglement-based CV QKD employing heterodyne detection is also possible \cite{PhysRevLett.93.170504,PhysRevLett.110.030502,PhysRevLett.114.070501,PhysRevLett.118.200501}; however, the entropic uncertainty relations are not applicable for proving security in such schemes \cite{PhysRevLett.110.030502,PhysRevLett.118.200501}, so we do not consider them here as our result relates to a modification of the entropic uncertainty relations.

In close analogy to time-frequency QKD, for a homodyne-based CV QKD protocol, Alice sorts her results into bins, yielding the POVMs for $X$ and $P$ quadrature measurements with elements \cite{Furrer2012}:
\begin{equation}
\mathbb{X}_{A}^{m}=\int_{x_{m}-\delta/2}^{x_{m}+\delta/2}dx|x\rangle\langle x|,\ \ \mathbb{P}_{A}^{k}=\int_{p_{k}-\delta /2}^{p_{k}+\delta /2}dp|p\rangle\langle p|
\end{equation}
where $|x\rangle$ and $|p\rangle$ are the eigenstates of the quadrature operators. If the constant bin width, $\delta$, can be maintained across the entire measurement range, then we should expect an expression for the maximum overlap just like Eq. (10), since that equation is derived for general continuous variables respecting Heisenberg-like uncertainty relations \cite{Rudnicki2012}.

Of course, as with time-frequency QKD, we cannot expect the same level of coarse-graining over the entire Hilbert space. Above some intensity level, the detectors will become saturated. The measurement operators characterizing saturation in the $X$ and $P$ measurements will be defined over semi-infinite ranges, meaning $c\approx 1$ \cite{Furrer2012}. The suggestion made in \cite{Furrer2012} for dealing with this issue is that one needs to trust the source, assume it has a low probability of saturating the detectors, and then incorporate that probability into the smoothing parameter for the entropic uncertainty relations resulting in a failure probability for the protocol. In \cite{Gehring2015}, the solution is to assume that the energy of the source is bounded, so that the probability for saturation can be estimated.

This issue has been discussed in \cite{Qin2015,Qin2018}, in the context of the Gaussian modulated coherent state protocol, a prepare-and-measure setting. They found that an eavesdropper can shift the mean of the distribution of results into the saturation regime, simultaneously lowering the variance of results, which causes Alice and Bob to overestimate the security of their key. Among countermeasures suggested, they discussed introducing a confidence interval for the results, and if too many results fall beyond the confidence interval, Alice and Bob ought to abort the protocol; however, the range of this confidence interval and the threshold probability for aborting the protocol were left open for future work. 

Our main result, Eq. (6), addresses this gap: our bound depends explicitly on the probabilities of saturating the detectors, $p_{X_{A}}^{\emptyset}$ and $p_{P_{A}}^{\emptyset}$, which can be measured  from the data without having to trust the source. This provides a way for Alice and Bob to monitor how many results are beyond their confidence interval, guarding against a saturation attack by Eve. Note that, by depending on $p_{X_{A}}^{\emptyset}$ and $p_{P_{A}}^{\emptyset}$, our bound additionally provides an implicit way to include the choice of the measurement range as an optimization parameter for the protocol.

Note the major difference between the problematic measurements in time-frequency and CV QKD using homodyne detection. In the former, losing a photon had the same consequence for security as falling outside the measurement range, since both resulted in the detector not clicking; the high loss in fiber optic channels compromised security after a short distance. In homodyne-based CV QKD, however, saturation is problematic, and luckily the fiber optic channel will not naturally introduce gain that will convert a low intensity signal into a saturating signal. Other than tampering, the main source for detector saturation is due to the tails of the Gaussian distributions in phase space from the two-mode squeezed vacuum states used for entanglement-based protocols. Luckily, these probabilities are vanishingly small. For example, in \cite{Gehring2015}, the range of measurement is [-61.6, 61.6] in units of vacuum noise, and using an anti-squeezing factor of 19.3 dB \cite{Mehmet2011}, this yields $p_{X_{A}}^{\emptyset} = p_{P_{A}}^{\emptyset} \approx 0$ to within machine precision. We should not expect the results with the new bound to differ from \cite{Gehring2015}, since Eq. (6) reduces to Eq. (2) when those probabilities are zero.

\section{Conclusion}
We have presented a modified entropic uncertainty relation to discount unwanted POVM elements that render the unmodified entropic uncertainty relation trivial. This is done at the cost of including the probability of the unwanted measurement in the bound. Our bound offers insight into the measurement range problem, which poses an issue for the characterization of entanglement in high-dimensional systems. We applied the bound to analyze entanglement-based time-frequency QKD, and found that, unlike previous results, we can now guarantee security; however, this is conditional on low loss and high detection efficiency within the measurement range. In a practical setting, this may only be achievable with some characterization of the source, like knowing the probability a vacuum state is emitted, weakening the completely untrusted source assumption of entanglement-based QKD. Under realistic conditions, the key becomes insecure at $\sim$2 km, mainly due to the high loss in fiber. Finally, we discussed our bound as it related to saturation attacks in CV QKD employing homodyne detection schemes. Through the bound's dependence on the probability of detector saturation, we provide a new quantitative way to guard against saturation attacks, without any assumptions about the source but with full characterization of the measurements.

\section*{Funding}
Natural Sciences and Engineering Research Council of Canada (NSERC) Canada Graduate Scholarship - Master's; Ontario Graduate Scholarship; Lachlan Gilchrist Fellowship Fund; US Office of Naval Research; Huawei Technologies
Canada Co., Ltd.

\section*{Acknowledgments}
J. E. B. is grateful for helpful discussions with Prof. Li Qian, Prof. Charles Ci Wen Lim, Dr. Bing Qi, Prof. Zheshen Zhang, Prof. Norbert L\"utkenhaus, and Aaron Goldberg, and particularly indebted to Ilan Tzitrin for discussions and providing feedback on an earlier version of this paper.

\bibliography{My_Collection}

\section*{Appendix 1: Basis-independent null measurements pose no problem for entropic uncertainty relations}
We show that if the null measurement is independent of the measurement type, then we can always reformulate an entropic uncertainty relation that does not depend on the null measurement operator common to both measurement types. However, it depends on new, effective POVMs related to the original POVMs. We show how those can be constructed.

Let $\mathcal{H}$ be the total Hilbert space. Suppose we have two measurements, $Z$ and $X$, characterized by
POVMs $Z=\{\mathbb{Z}_{n}\}_{n=0}^{N_{Z}}\cup\mathbb{N}$ and $X=\{\mathbb{X}_{m}\}_{m=0}^{N_{X}}\cup\mathbb{N}$,
where $\mathbb{N}=\mathbb{I}-\sum_{n=0}^{N_{Z}}\mathbb{Z}_{n}=\mathbb{I}-\sum_{m=0}^{N_{X}}\mathbb{X}_{m}=\mathbb{I}-\mathbb{M}$
represents the basis-independent null measurement. 

We can define $\mathcal{H}_\mathbb{N}$ ($\mathcal{H}_\mathbb{M}$) to be the space on which $\mathbb{N}$ ($\mathbb{M}$) has support. In general, $\mathcal{H}_\mathbb{N}$ and $\mathcal{H}_\mathbb{M}$ will have overlap, with $\mathcal{H}_\mathbb{N}\cup\mathcal{H}_\mathbb{M}=\mathcal{H}$.

With the definition of those spaces in hand, we can treat some operators in the problem as the direct sum of operators with support only on $\mathcal{H}_\mathbb{M}$, and of the zero operator on $\mathcal{H}/\mathcal{H}_\mathbb{M}$:
\begin{equation}
\begin{split}
\mathbb{M}=&\ (\mathbb{M})_{\mathcal{H}_\mathbb{M}}\oplus 0_{\mathcal{H}/\mathcal{H}_\mathbb{M}} \\
\mathbb{Z}_n=&\ (\mathbb{Z}_n)_{\mathcal{H}_\mathbb{M}}\oplus 0_{\mathcal{H}/\mathcal{H}_\mathbb{M}}\\
\mathbb{X}_m=&\ (\mathbb{X}_m)_{\mathcal{H}_\mathbb{M}}\oplus 0_{\mathcal{H}/\mathcal{H}_\mathbb{M}}
\end{split}
\end{equation}
where $(\cdot)_{\mathcal{H}_\mathbb{M}}$ denotes an operator with only support on $\mathcal{H}_\mathbb{M}$.

\textbf{Claim:} $[\sqrt{(\mathbb{M})_{\mathcal{H}_\mathbb{M}}}]^{-1}$ is well-defined.

\textbf{Proof:} As a POVM element, $\mathbb{M}$ is Hermitian and non-negative, so all its eigenvalues are real and non-negative. Moreover, when we consider the operator $(\mathbb{M})_{\mathcal{H}_\mathbb{M}}$, we have now restricted to only the eigenvalues that are positive. Thus, the square root of $(\mathbb{M})_{\mathcal{H}_\mathbb{M}}$ is well-defined, and invertible. $\blacksquare$

We can now define new operators:
\begin{equation}
\begin{split}
\tilde{\mathbb{Z}}_n=&\  \big[\sqrt{(\mathbb{M})_{\mathcal{H}_\mathbb{M}}}\big]^{-1}(\mathbb{Z}_n)_{\mathcal{H}_\mathbb{M}} \big[\sqrt{(\mathbb{M})_{\mathcal{H}_\mathbb{M}}}\big]^{-1}\\
\tilde{\mathbb{X}}_m=&\  \big[\sqrt{(\mathbb{M})_{\mathcal{H}_\mathbb{M}}}\big]^{-1}(\mathbb{X}_m)_{\mathcal{H}_\mathbb{M}} \big[\sqrt{(\mathbb{M})_{\mathcal{H}_\mathbb{M}}}\big]^{-1}.
\end{split}
\end{equation}

\textbf{Claim:} $\tilde{Z}=\{\tilde{\mathbb{Z}}_n\}_{n=0}^{N_Z}$ and $\tilde{X}=\{\tilde{\mathbb{X}}_m\}_{m=0}^{N_X}$ each form a POVM on $\mathcal{H}_\mathbb{M}$.

\textbf{Proof:} To show $\tilde{Z}$ is a POVM we need to show its elements are Hermitian and non-negative, and that they sum to $\mathbb{I}_\mathbb{M}$. Because $(\mathbb{Z}_n)_{\mathcal{H}_\mathbb{M}}$ and  $[\sqrt{(\mathbb{M})_{\mathcal{H}_\mathbb{M}}}]^{-1}$ are Hermitian and non-negative, then $\tilde{\mathbb{Z}}_n= \big[\sqrt{(\mathbb{M})_{\mathcal{H}_\mathbb{M}}}\big]^{-1}(\mathbb{Z}_n)_{\mathcal{H}_\mathbb{M}}\big[\sqrt{(\mathbb{M})_{\mathcal{H}_\mathbb{M}}}\big]^{-1}$ is also Hermitian and non-negative. Moreover, we can check:
\begin{equation}
\begin{split}
\sum_{n=0}^{N_Z}\tilde{\mathbb{Z}}_n=&\big[\sqrt{(\mathbb{M})_{\mathcal{H}_\mathbb{M}}}\big]^{-1}\sum_{n=0}^{N_Z}(\mathbb{Z}_n)_{\mathcal{H}_\mathbb{M}} \big[\sqrt{(\mathbb{M})_{\mathcal{H}_\mathbb{M}}}\big]^{-1}\\
=&\big[\sqrt{(\mathbb{M})_{\mathcal{H}_\mathbb{M}}}\big]^{-1}(\mathbb{M})_{\mathcal{H}_\mathbb{M}} \big[\sqrt{(\mathbb{M})_{\mathcal{H}_\mathbb{M}}}\big]^{-1}=\mathbb{I}_\mathbb{M}.
\end{split}
\end{equation}
The same process follows for $\tilde{X}$. $\blacksquare$

Now that we have two new effective POVMs,
we will use them to understand an equivalence between two different scenarios:

\textbf{Scenario 1 (a ``realistic'' model):} An untrusted source
outputs to Alice a state $\rho$ that could be entangled with Eve
and/or Bob. With probability $q$, Alice performs a Z measurement,
and with probability $(1-q)$ Alice performs an X measurement. Alice
knows what type of measurement she is performing. Both Z and X measurements
include the null measurement element, $\mathbb{N}$. Depending on
if she performs a Z measurement or an X measurement, Alice records
the result in separate registers, $R_{Z}$ or $R_{X}$, respectively.
After many iterations, Alice considers two probability distributions: $p(m|\rho,X,\lnot\mathbb{N})$ ($p(n|\rho,Z,\lnot\mathbb{N})$) is the distribution of measurement
results given she performed an X (Z) measurement on the incoming
state, $\rho$, and post-selected to discard the null measurement.
This distribution represents the state of the register, $R_{X}$ ($R_{Z}$),
having discarded null measurements. The POVMs characterizing the Z and X measurements
have been outlined in the previous section.

\textbf{Scenario 2 (a fictitious model):} An untrusted source outputs
the same state $\rho$, but before it is sent to Alice, Eve passes
it through a filter that blocks the state if it returns the null measurement,
and lets it pass if it does not return the null measurement. The state
Alice then receives is:
\begin{equation}
\rho'=\frac{\sqrt{\mathbb{M}}\rho\sqrt{\mathbb{M}}}{\text{Tr}(\rho\mathbb{M})}.
\end{equation}
Note that this state is only non-zero on $\mathcal{H}_\mathbb{M}$
because it filters out the portion of the state that has support on
$\mathcal{H}/\mathcal{H}_{\mathbb{M}}$.

Now Alice performs either one of the two new measurements we defined, $\tilde{Z}$ or $\tilde{X}$, which are related to, but different
from, $Z$ and $X$. Similarly to before, depending on if she performs a $\tilde{Z}$ measurement
or an $\tilde{X}$ measurement, Alice records the result in separate
registers, $\tilde{R}_{Z}$ or $\tilde{R}_{X}$, respectively. After
many iterations, Alice considers two probability distributions: $p(m|\rho',\tilde{X})$ ($p(n|\rho',\tilde{Z})$) is the distribution of measurement results
given that an $\tilde{X}$ ($\tilde{Z}$) measurement was performed on the incoming
state, $\rho'$. This distribution represents the state of the register,
$\tilde{R}_{X}$ ($\tilde{R}_{Z}$).

One can check that:
\begin{equation}
\begin{split}
p(m|\rho',\tilde{X})=&\ p(m|\rho,X,\lnot\mathbb{N})=\frac{\text{Tr}(\rho\mathbb{X}_m)}{\text{Tr}(\rho\mathbb{M})}\\
p(n|\rho',\tilde{Z})=&\ p(n|\rho,Z,\lnot\mathbb{N})=\frac{\text{Tr}(\rho\mathbb{Z}_n)}{\text{Tr}(\rho\mathbb{M})}.
\end{split}
\end{equation}
Thus, because Alice
cannot distinguish between these two distributions, we can assume
Eve can have control of when a null measurement outcome will occur.

We can now examine an entropic
uncertainty relation for Scenario 2. Note that an entropic uncertainty
relation for Scenario 1 would lead to the trivial lower bound of 0
on $H_{\text{min}}(Z|E)$ because of the overlap between the common null
measurement element in both POVM sets: $c(X,Z)=||\mathbb{N}||_{\infty}^{2}=1$
because the eigenvalues of $\mathbb{N}$ on $\mathcal{H}/\mathcal{H}_\mathbb{M}$ are
all 1.

Because $\tilde{X}$ and $\tilde{Z}$
are both POVMs on $\mathcal{H}_\mathbb{M}$, the only space on which $\rho'$ can have support, then:
\[
H_{\text{min}}(\tilde{Z}|E)_{\rho'}+H_{\text{max}}(\tilde{X}|B)_{\rho'}\geq-\log\max_{m,n}||\sqrt{\tilde{\mathbb{Z}}_{n}}\sqrt{\tilde{\mathbb{X}}_{m}}||^{2}.
\]
Hence, we have provided a proof that basis-independent
null measurements do not pose a problem for entanglement-based QKD
protocols.

Note that the same process cannot be done with basis-dependent null measurements since there would not necessarily be a common POVM element $\mathbb{N}$. Thus, defining the subspace, $\mathcal{H}_\mathbb{M}$, would not be possible, precluding the definition of new effective POVMs, $\tilde{Z}$ and $\tilde{X}$, and the reduction to an entropic uncertainty relation in terms of just those operators.

\section*{Appendix 2: Proof of main result, Eq. (6)}
We assume a tripartite state, $\rho_{ABE}$, and two POVMs on $\mathcal{H}_A$, $Z=\{\mathbb{Z}_{A}^{z}\}_{z=1}^{N_{Z}}\cup\{\mathbb{Z}_{A}^{\emptyset}\}$ and $X=\{\mathbb{X}_{A}^{x}\}_{x=1}^{N_{X}}\cup\{\mathbb{X}_{A}^{\emptyset}\}$. We begin by following the procedure from \cite{Coles2012}. Define $\rho_{ZZ'ABE}=V_{ZZ'}\rho_{ABE}V_{ZZ'}^{\dagger}$, where $V_{ZZ'}=\sum_{z=1}^{N_Z}|z\rangle_{Z}|z\rangle_{Z'}\sqrt{\mathbb{Z}_{A}^{z}}+|\emptyset\rangle_{Z}|\emptyset\rangle_{Z'}\sqrt{\mathbb{Z}_{A}^{\emptyset}}$ is an isometry that maps the input state to a state entangled with the Z register. As in \cite{Coles2012}, we start from the duality relation:
\begin{equation}
H_{\text{min}}(Z|E)+H_{\text{max}}(Z|Z'AB)\geq0.
\end{equation}

Based on the definition for conditional max entropy in Eq. (3), we see our task is to upper bound $\max_{\sigma_{Z'AB}}F(\rho_{ZZ'AB},\mathbb{I}_{Z}\otimes\sigma_{Z'AB})$. One of our objectives is to remove dependence on $\mathbb{Z}_{A}^{\emptyset}$, so we first prove a useful lemma.

\textbf{Lemma:} For a composite Hilbert space, $\mathcal{H}_C\otimes\mathcal{H}_D$, if $\sqrt{\Gamma_C}\sqrt{\ \mathbb{I}_C-\Gamma_C}=0$ for some positive operator $\Gamma_C$ on $\mathcal{H}_C$ such that $\mathbb{I}_C-\Gamma_C$ is also positive, then $F(\rho_{CD},\mathbb{I}_C\otimes\sigma_D)\leq F(\rho_{CD},\Gamma_C\otimes\sigma_D)+F(\rho_{CD},(\mathbb{I}_C-\Gamma_C)\otimes\sigma_D)$.

\textbf{Proof:} Consider the trace norm formulation of the fidelity, and recall the triangle inequality for norms. Thus,
\begin{equation}
F(\rho_{CD},\mathbb{I}_C\otimes\sigma_D) = ||\sqrt{\rho_{CD}}\sqrt{\mathbb{I}_C\otimes\sigma_D}||_{\text{Tr}}\leq ||\sqrt{\rho_{CD}}\sqrt{\Gamma_C\otimes\sigma_D}||_{\text{Tr}}+||\sqrt{\rho_{CD}}\sqrt{(\mathbb{I}_C-\Gamma_C)\otimes\sigma_D}||_{\text{Tr}}
\end{equation}
on condition we can write $\sqrt{\mathbb{I}_C\otimes\sigma_D}=\sqrt{\Gamma_C\otimes\sigma_D}+\sqrt{(\mathbb{I}_C-\Gamma_C)\otimes\sigma_D}$. Consider the square of both sides: 
\begin{equation}
\begin{split}
\mathbb{I}_C\otimes\sigma_D&=\mathbb{I}_C\otimes\sigma_D+\sqrt{\Gamma_C\otimes\sigma_D}\sqrt{(\mathbb{I}_C-\Gamma_C)\otimes\sigma_D}+\sqrt{(\mathbb{I}_C-\Gamma_C)\otimes\sigma_D}\sqrt{\Gamma_C\otimes\sigma_D}\\
&= \mathbb{I}_C\otimes\sigma_D+\sqrt{\Gamma_C}\sqrt{(\mathbb{I}_C-\Gamma_C)}\otimes\sigma_D+\sqrt{(\mathbb{I}_C-\Gamma_C)}\sqrt{\Gamma_C}\otimes\sigma_D.
\end{split}
\end{equation}
We see the equality holds if $\sqrt{\Gamma_C}\sqrt{\ \mathbb{I}_C-\Gamma_C}=0$. $\blacksquare$

Note that $\mathbb{I}_{Z}\otimes\sigma_{Z'AB}=|\emptyset\rangle\langle\emptyset|_Z\otimes\sigma_{Z'AB}+\sum_{z=1}^{N_Z}|z\rangle\langle z|_Z\otimes\sigma_{Z'AB}$, and $\sqrt{|\emptyset\rangle\langle\emptyset|_Z}\sqrt{\sum_{z=1}^{N_Z}|z\rangle\langle z|_Z}=0$, so we use the lemma to find:
\begin{equation}
F(\rho_{ZZ'AB},\mathbb{I}_{Z}\otimes\sigma_{Z'AB})\leq F(\rho_{ZZ'AB},|\emptyset\rangle\langle\emptyset|_{Z}\otimes\sigma_{Z'AB})+F(\rho_{ZZ'AB},\sum_{z=1}^{N_Z}|z\rangle\langle z|_{Z}\otimes\sigma_{Z'AB}).
\end{equation}
Using the data-processing inequality for fidelities \cite{Tomamichel2012a}, $F(\rho,\sigma)\leq F[\mathcal{E}(\rho),\mathcal{E}(\sigma)]$, where $\mathcal{E}(\cdot)$ is a trace-preserving, completely positive map, we find:
\begin{equation}
F(\rho_{ZZ'AB},|\emptyset\rangle\langle\emptyset|_{Z}\otimes\sigma_{Z'AB})\leq F(\rho_Z,|\emptyset\rangle\langle\emptyset|_{Z})=\sqrt{p_{Z_{A}}^{\emptyset}}.
\end{equation}

Next, we use the fact from \cite{Coles2012} that relative entropies are invariant under isometries. Since the max relative entropy is simply proportional to the logarithm of the fidelity, this means fidelity is also invariant under isometries. Following the process done in equation (6) of the supplementary material of \cite{Coles2012}, we get:
\begin{equation}
\begin{split}
F(\rho_{ZZ'AB},\sum_{z=1}^{N_Z}|z\rangle\langle z|_{Z}\otimes\sigma_{Z'AB})&\leq F(\rho_{AB},V_{ZZ'}^{\dagger}\sum_{z=1}^{N_Z}|z\rangle\langle z|_{Z}\otimes\sigma_{Z'AB}V_{ZZ'})\\ &= F\big[\rho_{AB},\text{Tr}_{Z'}(\sum_{z=1}^{N_Z}|z\rangle\langle z|\sqrt{\mathbb{Z}_{A}^{z}}\sigma_{Z'AB}\sqrt{\mathbb{Z}_{A}^{z}})\big].
\end{split}
\end{equation}
Note that $\text{Tr}_{Z'}(\sum_{z=1}^{N_Z}|z\rangle\langle z|\sqrt{\mathbb{Z}_{A}^{z}}\sigma_{Z'AB}\sqrt{\mathbb{Z}_{A}^{z}})$ may not be a normalized density matrix, but we will fix this later.

Now, we define the isometry associated with the X-type measurement, $V_{XX'}=\sum_{x=1}^{N_X}|x\rangle_{X}|x\rangle_{X'}\sqrt{\mathbb{X}_{A}^{x}}+|\emptyset\rangle_{X}|\emptyset\rangle_{X'}\sqrt{\mathbb{X}_{A}^{\emptyset}}$, use again the fact that fidelity is invariant under isometries, and again use the data-processing inequality to trace over subsystems $A$ and $X'$:
\begin{equation}
\begin{gathered}
F\big[\rho_{AB},\text{Tr}_{Z'}(\sum_{z=1}^{N_Z}|z\rangle\langle z|\sqrt{\mathbb{Z}_{A}^{z}}\sigma_{Z'AB}\sqrt{\mathbb{Z}_{A}^{z}})\big] \\
\leq F\bigg\{\rho_{XB},\sum_{x}|x\rangle\langle x|_{X}\otimes \text{Tr}_{Z'A}\bigg[\sum_{z=1}^{N_Z}\bigg(|z\rangle\langle z|_{Z'}\otimes\sqrt{\mathbb{Z}_{A}^{z}}\mathbb{X}_{A}^{x}\sqrt{\mathbb{Z}_{A}^{z}}\bigg)\sigma_{Z'AB}\bigg]\bigg\}
\end{gathered}
\end{equation}
where $\rho_{XB}=\sum_{x}|x\rangle\langle x|_X\otimes \text{Tr}_{A}\big(\rho_{AB}\mathbb{X}_{A}^{x}\big)$. We now note:
\begin{equation}
\begin{gathered}
F\bigg\{\rho_{XB},\sum_{x}|x\rangle\langle x|_{X}\otimes \text{Tr}_{Z'A}\bigg[\sum_{z=1}^{N_Z}\bigg(|z\rangle\langle z|_{Z'}\otimes\sqrt{\mathbb{Z}_{A}^{z}}\mathbb{X}_{A}^{x}\sqrt{\mathbb{Z}_{A}^{z}}\bigg)\sigma_{Z'AB}\bigg]\bigg\} \\
=F\bigg\{|\emptyset\rangle\langle\emptyset|_X\otimes \text{Tr}_{A}\big(\rho_{AB}\mathbb{X}_{A}^{\emptyset}\big),|\emptyset\rangle\langle\emptyset|_X\otimes \text{Tr}_{Z'A}\bigg[\sum_{z=1}^{N_Z}\bigg(|z\rangle\langle z|_{Z'}\otimes\sqrt{\mathbb{Z}_{A}^{z}}\mathbb{X}_{A}^{\emptyset}\sqrt{\mathbb{Z}_{A}^{z}}\bigg)\sigma_{Z'AB}\bigg]\bigg\}\\
+F\bigg\{\sum_{x=1}^{N_X}|x\rangle\langle x|_X\otimes \text{Tr}_{A}\big(\rho_{AB}\mathbb{X}_{A}^{x}\big),\sum_{x=1}^{N_X}|x\rangle\langle x|_X\otimes \text{Tr}_{Z'A}\bigg[\sum_{z=1}^{N_Z}\bigg(|z\rangle\langle z|_{Z'}\otimes\sqrt{\mathbb{Z}_{A}^{z}}\mathbb{X}_{A}^{x}\sqrt{\mathbb{Z}_{A}^{z}}\bigg)\sigma_{Z'AB}\bigg]\bigg\}.
\end{gathered}
\end{equation}
Applying the data-processing inequality to the first term to trace over $B$:
\begin{equation}
\begin{gathered}
F\bigg\{|\emptyset\rangle\langle\emptyset|_X\otimes \text{Tr}_{A}\big(\rho_{AB}\mathbb{X}_{A}^{\emptyset}\big),|\emptyset\rangle\langle\emptyset|_X\otimes \text{Tr}_{Z'A}\bigg[\sum_{z=1}^{N_Z}\bigg(|z\rangle\langle z|_{Z'}\otimes\sqrt{\mathbb{Z}_{A}^{z}}\mathbb{X}_{A}^{\emptyset}\sqrt{\mathbb{Z}_{A}^{z}}\bigg)\sigma_{Z'AB}\bigg]\bigg\}\\
\leq F(p_{X_A}^{\emptyset}|\emptyset\rangle\langle\emptyset|_X,|\emptyset\rangle\langle\emptyset|_{X})= \sqrt{p_{X_A}^{\emptyset}}.
\end{gathered}
\end{equation}
Additionally, we can define $\rho_{X^<B}=\sum_{x=1}^{N_X}|x\rangle\langle x|_X\otimes \text{Tr}_{A}\big(\rho_{AB}\mathbb{X}_{A}^{x}\big)/(1-p_{X_A}^{\emptyset})$ to be a normalized density operator, so that:
\begin{equation}
\begin{gathered}
F\bigg\{\sum_{x=1}^{N_X}|x\rangle\langle x|_X\otimes \text{Tr}_{A}\big(\rho_{AB}\mathbb{X}_{A}^{x}\big),\sum_{x=1}^{N_X}|x\rangle\langle x|_X\otimes \text{Tr}_{Z'A}\bigg[\sum_{z=1}^{N_Z}\bigg(|z\rangle\langle z|_{Z'}\otimes\sqrt{\mathbb{Z}_{A}^{z}}\mathbb{X}_{A}^{x}\sqrt{\mathbb{Z}_{A}^{z}}\bigg)\sigma_{Z'AB}\bigg]\bigg\} \\
= \sqrt{1-p_{X_A}^\emptyset}F\bigg\{\rho_{X^<B},\sum_{x=1}^{N_X}|x\rangle\langle x|_X\otimes \text{Tr}_{Z'A}\bigg[\sum_{z=1}^{N_Z}\bigg(|z\rangle\langle z|_{Z'}\otimes\sqrt{\mathbb{Z}_{A}^{z}}\mathbb{X}_{A}^{x}\sqrt{\mathbb{Z}_{A}^{z}}\bigg)\sigma_{Z'AB}\bigg]\bigg\}.
\end{gathered}
\end{equation}

Finally, we know from \cite{Coles2012} that if $\tilde{\sigma}\geq\sigma$, then $F(\rho,\tilde{\sigma})\geq F(\rho, \sigma)$, so:
\begin{equation}
\begin{gathered}
F\bigg\{\rho_{X^<B},\sum_{x=1}^{N_X}|x\rangle\langle x|_X\otimes \text{Tr}_{Z'A}\bigg[\sum_{z=1}^{N_Z}\bigg(|z\rangle\langle z|_{Z'}\otimes\sqrt{\mathbb{Z}_{A}^{z}}\mathbb{X}_{A}^{x}\sqrt{\mathbb{Z}_{A}^{z}}\bigg)\sigma_{Z'AB}\bigg]\bigg\} \\
\leq \max_{(x,z)\neq\emptyset}||\sqrt{\mathbb{Z}_{A}^{z}}\sqrt{\mathbb{X}_{A}^{x}}||_\infty  F\bigg\{\rho_{X^<B},\mathbb{I}_{X^<}\otimes \text{Tr}_{Z'A}\bigg[\sum_{z=1}^{N_Z}(|z\rangle\langle z|_{Z'})\sigma_{Z'AB}\bigg]\bigg\}.
\end{gathered}
\end{equation}
Applying this property again for $\sigma_B \geq \text{Tr}_{Z'A}[\sum_{z=1}^{N_Z}(|z\rangle\langle z|_{Z'})\sigma_{Z'AB}]$, we get:
\begin{equation}
\begin{split}
\max_{\sigma_{Z'AB}}F\bigg\{\rho_{X^<B},\mathbb{I}_{X^<}\otimes \text{Tr}_{Z'A}\bigg[\sum_{z=1}^{N_Z}(|z\rangle\langle z|_{Z'})\sigma_{Z'AB}\bigg]\bigg\} &\leq \max_{\sigma_B}F(\rho_{X^<B},\mathbb{I}_{X^<}\otimes \sigma_{B}) \\ &=\sqrt{2}^{H_{\text{max}}(X_A^<|B)}.
\end{split}
\end{equation}

Putting together Eq. (18), and (21)-(29), we get the result in Eq. (6). $\blacksquare$

\section*{Appendix 3: Smooth version of main result}
Smooth min- and max- entropies are useful quantities for incorporating finite key size effects \cite{Tomamichel2012,Tomamichel2011,Tomamichel2012a,Konig2009,Niu2016,Coles2017}. $\varepsilon$-smooth conditional min- and max-entropies are defined as:
\begin{equation}
H_{\text{min}}^{\varepsilon}(A|B)_{\rho}=\max_{\rho'\in\mathcal{B}^{\varepsilon}(\rho)}H_{\text{min}}(A|B)_{\rho'},\  H_{\text{max}}^{\varepsilon}(A|B)_{\rho}=\min_{\rho'\in\mathcal{B}^{\varepsilon}(\rho)}H_{\text{max}}(A|B)_{\rho'}
\end{equation}
where $\mathcal{B}^{\varepsilon}(\rho)=\{\rho'|\ \tfrac{1}{2}||\rho-\rho'||_{\text{Tr}}\leq\varepsilon\}$ denotes the set of operators within an $\varepsilon$-distance of $\rho$.

Proceeding similarly to the proof from \cite{Tomamichel2011}, we first note that for some $\tau\in\mathcal{B}^{\varepsilon}(\rho)$, $H_{\text{max}}^{\varepsilon}(X_A^<|B)_{\rho}=H_{\text{max}}(X_A^<|B)_{\tau}$. Thus, we can write down our main result, Eq. (6) for that state:
\begin{equation}
H_{\text{min}}(Z_{A}|E)_\tau\geq-2\log\left[\sqrt{p_{Z_{A}}^{\emptyset}(\tau)}+\sqrt{p_{X_{A}}^{\emptyset}(\tau)}+\sqrt{1-p_{X_{A}}^{\emptyset}(\tau)}\sqrt{c^{<}(X,Z)}\left(\sqrt{2}^{H_{\text{max}}^{\varepsilon}(X_A^<|B)_{\rho}}\right)\right]
\end{equation}
where the $p_{i}^{\emptyset}(\tau)$ denote the null measurement probabilities given the state $\tau$.

Next, we would like to express $p_{i}^{\emptyset}(\tau)$ in terms of $p_{i}^{\emptyset}(\rho)$ and $\varepsilon$ to determine how much the probabilities from the two states can differ. Using $1-F(\rho,\tau)\leq \tfrac{1}{2}||\rho-\tau||_{\text{Tr}}$ \cite{Konig2009}, and the data-processing inequality \cite{Tomamichel2012a}, we find:
\begin{equation}
f_-[p_{i}^{\emptyset}(\rho),\varepsilon]\leq p_{i}^{\emptyset}(\tau)\leq f_+[p_{i}^{\emptyset}(\rho),\varepsilon]
\end{equation}
where
\begin{equation}
f_\pm[p_{i}^{\emptyset}(\rho),\varepsilon]=2\varepsilon+p_{i}^{\emptyset}(\rho)+2p_{i}^{\emptyset}(\rho)\varepsilon^2-4p_{i}^{\emptyset}(\rho)\varepsilon-\varepsilon^2\pm2(1-\varepsilon)\sqrt{p_{i}^{\emptyset}(\rho)\varepsilon[1-p_{i}^{\emptyset}(\rho)][2-\varepsilon]}.
\end{equation}

Finally, following \cite{Tomamichel2011}, knowing that $H_{\text{min}}^{\varepsilon}(Z_A|E)_{\rho}=\max_{\rho'\in\mathcal{B}^{\varepsilon}(\rho)}H_{\text{min}}(Z_A|E)_{\rho'}\geq H_{\text{min}}(Z_A|E)_\tau$, we get the smooth version of Eq. (6):
\begin{equation}
\begin{split}
H_{\text{min}}^{\varepsilon}(Z_{A}|E)_\rho\geq-2\log\bigg[&\sqrt{f_+[p_{Z_A}^{\emptyset}(\rho),\varepsilon]}+\sqrt{f_+[p_{X_A}^{\emptyset}(\rho),\varepsilon]}+\\
&\sqrt{1-f_-[p_{X_A}^{\emptyset}(\rho),\varepsilon]}\sqrt{c^{<}(X,Z)}\bigg(\sqrt{2}^{H_{\text{max}}^{\varepsilon}(X_A^<|B)_{\rho}}\bigg)\bigg].
\end{split}
\end{equation}

\end{document}